\documentclass{jpsj2}

\def\runtitle{
Theoretical Description of Nearly Discontinuous Transition in 
Superconductors with Paramagnetic Depairing}
\def\runauthor{Hiroto \textsc{Adachi}, Shigeru \textsc{Koikegami},and Ryusuke \textsc{Ikeda}
}

\title{%
Theoretical Description of Nearly Discontinuous Transition in 
Superconductors with Paramagnetic Depairing}

\author{%
Hiroto \textsc{Adachi}\thanks{ada@scphys.kyoto-u.ac.jp}, Shigeru \textsc{Koikegami}$^1$, 
and Ryusuke \textsc{Ikeda}
}

\inst{%
Department of Physics, Kyoto University, Kyoto 606-8502, Japan\\
$^1$Nanoelectronics Research Institute, AIST Tsukuba Central 2, 305-8568, Japan}

\recdate{\today}

\abst{%
Based on a theoretical argument and Monte Carlo simulations of 
a Ginzburg-Landau model derived microscopically, it is argued that, 
in type-II superconductors where {\it both} the paramagnetic 
{\it and} orbital depairings are important, a strong first order transition 
(FOT) at $H_{c2}$ expected in the mean field (MF) approximation never occurs 
in real systems and changes due to the fluctuation into a crossover. 
The present result explains why a {\it nearly} discontinuous crossover 
at $H_{c2}$ with {\it no} intrinsic hysteresis is observed only in a clean 
superconducting material with a singlet pairing and a high condensation 
energy such as CeCoIn$_5$. 
}

\kword{%
paramagnetic and orbital depairings, upper critical field, 
CeCoIn$_5$, vortex phase diagram
}

\begin{document}
\sloppy
\maketitle

   The phase diagram of a type-II superconductor in nonzero magnetic fields 
($H \neq 0$) has been studied in relation to high $T_c$ cuprates 
in fields of the tesla range and beyond the mean field (MF) approximation 
completely neglecting the superconducting (SC) fluctuation effect~\cite{Brandt,RIr}. 
Even in a material with a spin-singlet electron-pairing, 
the spin (i.e., Pauli paramagnetic) depairing effect can be 
neglected in relatively low fields, and the transition at $H_{c2}(T)$-line 
in the MF approximation for this case is of second order. 
On the other hand, the phase diagram in cases where {\it both} 
the Pauli paramagnetic and orbital depairings are important 
were examined so far only within the MF approximation \cite{Mineev,Houzet}. 
Since the MF transition in this case has been expected to become first order 
at low enough temperatures \cite{Mineev,Houzet}, it is valuable to clarify 
whether this MF scenario is also, as well as 
in the low field case \cite{Brandt,RIr}, drastically changed by including 
the SC fluctuation. 

 In this paper, we first examine the MF approximation of a quasi two-dimensional (2D) 
superconductor under a field {\it perpendicular} to the SC layers 
(${\mib H} \parallel c$) by treating the 
orbital- and spin-depairing effects on an equal footing and obtain a first 
order transition in the MF approximation (MF-FOT) at $H_{c2}(T)$ and 
in high fields. Then, based on a physical picture on the high $H$ 
fluctuation and our Monte Carlo simulation results on the derived 
Ginzburg-Landau (GL) model, we argue that any nonvanishing SC fluctuation, 
as well as in the low $H$ case \cite{Brandt,RIr}, changes this MF-FOT 
into a crossover and point out a consistency between the present theory 
and experimental results, such as the apparently discontinuous behavior 
\cite{Izawa,Tayama,Murphy,Bianchi} in high fields with {\it no} intrinsic 
hysteresis seen 
in CeCoIn$_5$ with unusually large condensation energy 
{\it and} the absence of such a sharp behavior in other 
materials \cite{ISSP,Bi,Naughton} showing large paramagnetic effects and 
having lower condensation energies. 

 Let us first briefly explain our microscopic derivation of a Ginzburg-Landau 
 (GL) functional in the case with spin depairing effect by starting with a 
 quasi 2D BCS hamiltonian. An attractive interaction leading to 
 the $d_{x^2-y^2}$-pairing state will be assumed. 
 For ${\mib H} \parallel c$ case, the following GL model will be derived 
microscopically for a layered system:
\begin{eqnarray}\label{eq:GL1}
{\cal F} &=& N(0) \int d^2{r}_\perp \Bigg[
      \sum_{q_z} \sum_N a_N(q_z^2) |\tilde{\Delta}_{q_z}^{(N)}({\mib r}_\perp)|^2 
\hspace{2cm}\nonumber \\ 
&+&  \sum_j
      \Big(  \frac{V_4}{2} |\Delta_j^{(0)}({\mib r}_\perp)|^4 + 
      \frac{V_6}{3} |\Delta_j^{(0)}({\mib r}_\perp)|^6    \Big) 
      \Bigg], 
\end{eqnarray}
where $\Delta_j^{(N)}({\mib r}_\perp) 
= N_{\rm layer}^{-1/2} \sum_{q_z} \tilde{\Delta}_{q_z}^{(N)}({\mib r}_\perp) 
e^{{\rm i}q_z s j}$ denotes the pair-field (SC order parameter) 
on the $j$-th SC layer projected onto the Nth Landau level (LL), $N(0)$ is 
a 2D density of state at the 
Fermi surface, $N_{\rm layer}$ is the number of SC layers, and $s$ is 
the layer spacing. Below, the higher LL modes with $N \geq 2$ will be neglected. 
In eq.(\ref{eq:GL1}), the 4th and 6th order terms were represented within the 
lowest LL approximation and in a spatially local form. 
The neglect of this non-locality is justified because the energy scale 
associated with the MF transition (i.e., the condensation energy) 
is much larger than that on 
a phenomenon arising from this non-locality such as a structural transition 
between the rhombic and square vortex lattices. 
Then, assuming, for brevity, a cylindrical Fermi surface and treating effects 
of the magnetic field quasi-classically, 
the coefficient $a_N(q_z^2)$ of eq.(\ref{eq:GL1}) in pure case 
($\tau^{-1}=0$) becomes 
\begin{eqnarray}
a_N (q_z^2) &=& \ln(T/T_{c0}) -2 \pi T \sum_{\varepsilon > 0}
       \left( D_N(2 \varepsilon) - 1/\varepsilon \right),\label{eq:aN}\\
D_N(2 \varepsilon) &=& 2\int_{0}^{\infty} d \rho
 e^{-2\varepsilon \rho - \big( \frac{\rho}{2\tau_H} \big)^2} 
 {\cal L}_N \Big( 2(\frac{\rho}{2\tau_H})^2 \Big) \nonumber \\
&& \quad \times \cos(2 \mu_0 H \rho) {\cal J}_0 \Big(2J\sin(\frac{q_z s}{2})\rho\Big). 
\label{eq:DN}
\end{eqnarray}
where $\varepsilon=2 \pi (n+1/2)T$ is a Matsubara frequency, 
$T_{c0}$ the transition temperature in $H=0$, $\tau_H=r_H/v_F$, 
$v_F$ the Fermi velocity, $r_H=\sqrt{\phi_0/2 \pi H}$ the magnetic length, 
$J$ the interlayer hopping, 
$\mu_0 H$ the Zeeman energy, and ${\cal L}_N$ and ${\cal J}_N$ 
are the Nth order Laguerre and Bessel functions, respectively. In the impure case 
with a finite elastic scattering rate $\tau^{-1}$, the impurity-ladder 
vertex correction can be neglected in non $s$-wave pairing cases 
if $\tau_H \ll \tau$ \cite{Mineev,Ada}. Then, the corresponding expression to 
eq.(\ref{eq:aN}) is given by replacing 
$D_N(2 \varepsilon)$ by $D_N(2 \varepsilon+\tau^{-1})$. 

\begin{figure}[t]
\begin{center}
\scalebox{0.5}[0.5]{\includegraphics{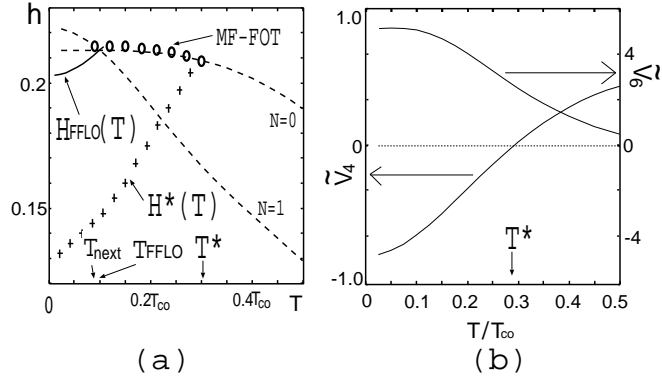}}
\end{center}
\caption{(a) Mean field $H$-$T$ phase diagram obtained from computations 
in high $H$ and low $T$ region. The magnetic field is measured in the reduced 
variable $h=H/H_{c2}^{\rm orbit}$. (b) $T$-dependences of ${\tilde V}_4=V_4/\tau_H^2$ 
and ${\tilde V}_6=V_6/\tau_H^4$ computed on $H_{c2}(T)$. In both (a) and (b), 
$(2 \pi \tau T_{c0})^{-1}=0.025$ was commonly used. See 
the text for other details.}
\label{fig:Hc2} \end{figure}

 The corresponding calculation of the coefficients $V_4$ and $V_6$ is more 
involved and requires multiple numerical 
integrals. The orbital depairing always introduces the field-induced vortices, 
which may change drastically the phase diagram in $H \neq 0$, and 
hence, will be included nonperturbatively in contrast to 
ref.\cite{Houzet}. Further, the paramagnetic effect becoming important upon 
cooling is sensitive to the impurity scattering. 
Hence, 
both the two depairing effects and the impurity scattering should be treated 
on 
an equal footing. For this reason, we have examined the coefficients $V_4$ 
and $V_6$ 
numerically by, as well as in eq.(\ref{eq:aN}), applying the parameter 
($\rho$ in 
eq.(\ref{eq:DN})) integral representation and neglecting the vertex 
corrections. 
Details of their derivation will be described elsewhere \cite{Ada}. 
Consequently, as shown in Fig.\ref{fig:Hc2}, 
the coefficient $V_4$ ($V_6$) defined in $N=0$ becomes negative (positive) 
in $H > H^*(T)$. 

In Fig.\ref{fig:Hc2} (a), 
we give a MF phase diagram in high $H$ following from 
calculations of the GL coefficients mentioned above in a slightly impure 
($(2 \pi \tau T_{c0})^{-1}=0.025$) case. 
There, the paramagnetic parameter $\mu_0 H_{c2}^{\rm orbit}/2 \pi T_{c0}$ 
corresponding to the Maki parameter is chosen as 0.8, 
where $H_{c2}^{\rm orbit}=0.56 \phi_0/(2 \pi \xi_0^2)$ is the 2D orbital 
limiting field in pure case, and $\xi_0=v_F/2\pi T_{c0}$. 
The dashed curves express the lines defined by $a_0(0)=0$ 
and $a_1(0)=0$, and, in a range where the MF transition is of second order, 
the higher field portion of these two curves plays the role of the 
$H_{c2}(T)$-line. Above the curve denoted by the plus symbols (i.e., $H > H^*(T)$), 
however, the coefficient $V_4$ in $N=0$ is negative, making the MF 
transition in $N=0$ a discontinuous one (i.e., FOT) which occurs on the 
portion of $H_{c2}(T)$ denoted by the open circles in $T_{\rm next} < 
T < T^*$. 
On the other hand, the MF transition in $N=1$ is found to remain continuous 
\cite{Ada}, although the $N=1$ region \cite{Shima} 
with $a_0(0) > a_1(0)$ is quite narrow. 
We have also estimated a structural transition curve $H_{\rm FFLO}(T)$ 
between the ordinary straight vortex solid and a 
FFLO-like helical vortex solid with a modulation 
$\Delta^{(0)}_j \sim e^{{\rm i}q_z sj}$ along ${\mib H}$ as the line where 
$\partial a_0(q_z^2)/\partial q_z^2|_{q_z=0}=0$. 
Since $\partial^2 a_0(q_z^2)/\partial (q_z^2)^2|_{q_z=0} >0$, 
and the spin depairing inducing this transition is enhanced with increasing $H$ 
and decreasing $T$, a {\it second order} structural transition at 
$H_{\rm FFLO}(T)$ {\it decreasing} upon cooling is expected at least near $H_{c2}$ 
to occur, 
although the range in which this $N=0$ state is realized seems to be quite 
narrow in this ${\mib H} \parallel c$ case. Actually, in contrast to $T^*$, 
the $H_{\rm FFLO}(T)$ curve and the $N=1$ vortex state region are very sensitive 
to the impurity strength, and we find that, for $(2 \pi T_{c0} \tau)^{-1}=0.05$, 
$T^* \simeq 0.23 T_{c0}$, while the $H_{\rm FFLO}(T)$ and the $N=1$ region 
are not present any longer. The main result $T^* > T_{\rm FFLO}$ 
in our MF calculation is satisfied, at least in ${\mib H} \parallel c$, 
irrespective of the impurity strength \cite{Ada} and should be compared 
with results in ref.\cite{Houzet}. 

 Now, let us turn to discussing the genuine high $H$ phase diagram of 
 superconductors with paramagnetic depairing. Hereafter, we will 
focus on the region $H^*(T) < H < H_{\rm FFLO}(T)$ 
in Fig.\ref{fig:Hc2}(a) and use the following GL functional on the basis 
of the MF results shown above: 
\begin{eqnarray} \label{eq:GL2}
\frac{{\cal F}}{T} 
&=& \sum_j \int d^2 r_{\perp} 
\Big\{ \alpha |\Psi_j|^2 + \gamma |\Psi_j - \Psi_{j+1}|^2 \nonumber \\
&& \hspace{1cm}      - \frac{|\beta(T)|}{2}|\Psi_j|^4+\frac{1}{3} |\Psi_j|^6 \Big\}. 
\end{eqnarray}  
Here, $\gamma > 0$, $\alpha \simeq \alpha_0(H-H_0)/H_0$ with a $T$-dependent 
constant $\alpha_0$, and $H_0$ is the field value satisfying $a_0(0)=0$. 
Further, the length and the pair-field were rescaled in the manner 
${\mib r_{\perp}}/r_H \to {\mib r_{\perp}}$ and 
$(T/ \, V_6 \, r_H^2 N(0))^{-1/6}\Delta_j^{(0)} \to \Psi_j$. Consequently, the 
2D fluctuation strength is measured by 
$|\beta(T)|^{-1} = |V_4|^{-1} (V_6^2 T/ N(0) r_H^2)^{1/3}$.
Any term expressing a vortex pinning effect induced by the 
electronic impurity scattering was neglected. 

   To explain why the MF-FOT should not occur in real systems with any 
nonvanishing fluctuation, it is convenient to start from 
a description deep in the ordered phase. 
First, since an inclusion of the orbital depairing requires the presence of 
field-induced vortices, a low energy excitation in the ordered phase is 
inevitably an elastic mode of a vortex solid. 
It is quite clear that the form of the elastic energy and the relation 
between the phase fluctuation and the vortex displacement \cite{Moore} 
are unaffected by a difference in the form of nonlinear terms of the GL model. 
As in the case with no paramagnetic depairing \cite{Moore}, 
the phase fluctuation is marginally relevant even in 3D case, 
and the main rigidity of the phase fluctuation is the shear modulus 
of the underlying vortex solid. 
Hence, if the melting of the vortex solid occurs at $H_m$ below $H_{c2}$, 
the vortex liquid just above $H_m$ with vanishing shear modulus has no long 
range 
orders and hence, should be continuously connected with the normal phase above 
$H_{c2}$ \cite{RIr}. In this case, the MF-FOT at $H_{c2}$ should not occur. 
Further, to understand this from another point of view, let us note that 
the quasi 2D SC order parameter in the lowest LL 
has the form~\cite{Tesanovic} 
\begin{equation}
\Psi(u, z) = {\cal A}(z) e^{-y^2/(2 r_H^2)} \prod_{i=0}^{N_s-1}
(u - u_i(z)), 
\end{equation} 
where $u=x+{\rm i}y$, $z=js$, $u_i(z)$ 
is the complex coordinate perpendicular to 
${\mib H}$ of the $i$-th vortex, and a Landau gauge was assumed for the 
external gauge field. 
Since the vortex positions are highly disordered above $H_m$, 
the fluctuation effect 
above $H_m$ is essentially described only by the amplitude ${\cal A}(z)$. 
However, ${\cal A}(z)$ depends only on $z$ {\it irrespective of} the form of 
nonlinear terms in the GL model. That is, since the amplitude 
fluctuation itself has a reduced dimensionality and is 1D-like in 3D 
systems \cite{RIr,Ikeda3} even in the present case, the amplitude fluctuation 
plays a role of pushing the vortex lattice freezing field down to a lower 
field, and the MF-FOT should not be realized in real 3D systems. 

\begin{figure}[t]
\begin{center}
\scalebox{0.4}[0.4]{\includegraphics{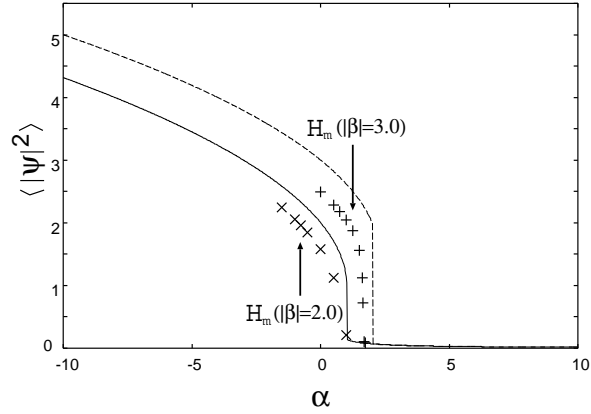}}
\end{center}
\caption{Numerical data (symbols) of $\langle|\Psi|^2 \rangle$ in the case 
composed of four layers. The corresponding results (curves) in the Hartree 
approximation \cite{Ikeda3} are also given. The cross symbols and solid curve 
(plus symbols and dotted curve) are results in $|\beta|=2.0$ ($3.0$). 
In $|\beta|=3.0$, the $H_{c2}(T)$ corresponds to $\alpha=1.7$ and lies above $H_m$. 
The used parameter values are given in the text on the final page.}
\label{fig:psi2} 
\end{figure}
 Bearing such a theoretical prospect in mind, we have performed Monte Carlo 
simulations on the model of eq.(\ref{eq:GL2}), and its 2D version using 
material 
parameter values useful for comparisons with observations in CeCoIn$_5$.
 Our simulation method follows that of ref.\cite{Kato} under the lateral 
boundary condition $L_x/L_y=2N_x/\sqrt{3}N_y$ with $(N_x,N_y)=(6,6)$ 
and a periodic boundary condition across the layers. Hereafter, 
the key quantities in the present case 
 within $N=0$ are $- \langle |\Psi|^2 \rangle$ and the structure factor 
defined \cite{Ikeda3} as the correlation function 
of $|\Psi({\mib r}_\perp)|^2$. 
The former is equivalent to the magnetization or the fluctuation entropy, 
while the latter is a measure of the vortex positional order. 
\begin{figure}[ht]
\begin{center}
\scalebox{0.4}[0.4]{\includegraphics{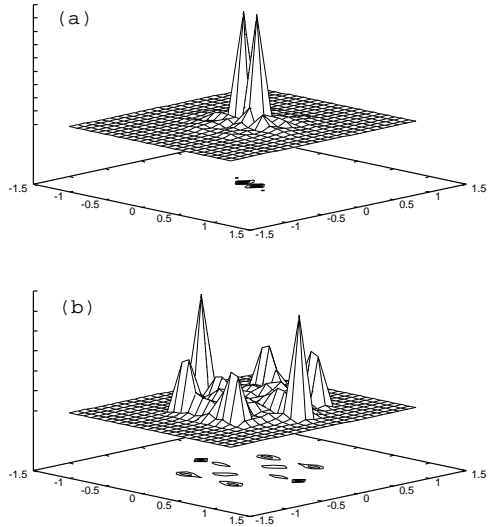}}
\end{center}
\caption{Structure factor defined from the correlation function of 
$|\Psi({\mib r})|^2$ for $|\beta|=3.0$ at (a)$\alpha=1.65$,(b) $\alpha=1.25$.}
\label{fig:Sq} 
\end{figure}

  Computed results of $\alpha$ dependences of $\langle |\Psi|^2 \rangle$ in 
a layered system consisting of four layers are shown in Fig.\ref{fig:psi2} for 
$|\beta|=2.0$ and $3.0$. Clearly, the nearly discontinuous behavior at $H_{c2}$ 
is smeared out as the 
fluctuation is enhanced (i.e., $|\beta|$ is reduced). 
For comparison, the corresponding results in the Hartree approximation~\cite{Ikeda3} 
are also drawn as the solid and dashed curves. 
Note that a change of $\langle |\Psi|^2 \rangle$ at $H_{c2}$ is extremely 
steep 
even in this approximation where {\it no} transition occurs \cite{Ikeda3}.
 Clearly, the presence of a genuine FOT cannot be argued merely through a 
steep growth of $\langle|\Psi|^2\rangle$ in a real system with weak SC fluctuation. 
Next, let us examine whether the melting transition point 
coincides or not with 
the MF transition field $H_{c2}(T)$.
 The melting transition is widely believed to be a {\it weak} FOT, 
and this should be found~\cite{Kato} in Monte Carlo simulations 
as a tiny discontinuity in thermodynamic quantities. 
Unfortunately, due primarily to numerical difficulties, 
our simulation is restricted to too small systems to observe 
such a discontinuity. 
For our purpose, however, it is sufficient to estimate the melting temperature 
by finding where the Bragg peaks of the vortex lattice disappear. 
Fig.\ref{fig:Sq} shows snapshots of the structure factor on 
$|\Psi|^2$ when $|\beta|=3.0$. No vortex positional ordering is seen 
at $H_{c2}$. As indicated in Fig.\ref{fig:psi2}, 
nearly sharp Bragg peaks appear at $H_m$ below $H_{c2}$, while most of the 
entropy have been lost rather near $H_{c2}$ above it. 
The two field (or temperature) scales, 
one characterizing the steep growth of $\langle |\Psi|^2 \rangle$ 
and another corresponding to the sudden growth of vortex positional 
ordering, are clearly distinguished. 

Quite similar results to those in Fig.\ref{fig:psi2} and \ref{fig:Sq} 
were also obtained in 2D case 
where the system size dependence of $H_m$ is more easily examined. As expected 
from a commensurability between a MF solution and the lateral periodic 
boundary 
condition, it was verified through the data in system sizes (6,4) and (8,6) 
with $L_x/L_y=\sqrt{3}N_x/2N_y$ that $H_m$ is lowered with increasing the system size.
 This means that the relation $H_m < H_{c2}$ found above holds in simulations 
for larger sizes. 

 Recently, a large apparent 
discontinuity was observed in transport and thermodynamic 
data \cite{Tayama,Murphy,Bianchi} in a heavy fermion material CeCoIn$_5$ 
in high fields, where a strong paramagnetic depairing is expected even 
in ${\mib H} \parallel c$ \cite{Tayama,SI}, and was identified with the MF-FOT. 
A hysteresis seen in the magnetization data was interpreted as being 
due to a genuine FOT at $H_{c2}$. 
However, the specific heat in ${\mib H} \parallel c$ \cite{Bianchi} and 
resistivity data \cite{Murphy} have shown {\it no} measurable hysteresis 
in contrast to observations \cite{Suzuki} in systems with no orbital depairing, 
although a stronger genuine FOT should be accompanied by a larger hysteresis 
reflecting a larger condensation energy. Further, a magnetic hysteresis 
in type-II superconductors is usually a consequence of vortex pinning \cite{03}.
Actually, a magnetization measurement in CeCo$_{1-x}$Rh$_x$In$_5$ 
has shown \cite{ISSP} that the discontinuity of magnetization disappears 
with increasing the Rh-substitution, while the hysteresis remains unaffected, 
implying that the observed \cite{Tayama,Murphy} magnetic hysteresis 
is {\it not} 
a direct consequence of an intrinsic SC transition. 
Further, the Rh-substitution has resulted in a decrease of condensation 
energy \cite{ISSP} which is usually equivalent to an enhancement of SC 
fluctuation. Together with the absence of a nearly discontinuous behavior 
in other systems \cite{Bi,Naughton} suggestive of a large spin depairing 
and with {\it strong} fluctuation, these observations in CeCoIn$_5$ are 
consistent with the 
absence of a genuine FOT stressed in this paper. 

Bearing these observations in Ce-compounds in mind, 
we have chosen parameter values in Fig.\ref{fig:psi2} and \ref{fig:Sq} as 
follows. 
The $N(0)$-value was taken from the heat capacity data 
in $H=0$ \cite{SI}. The Maki parameter value used in Fig.\ref{fig:Hc2} leading 
to 
$T^*/T_{c0} \simeq 0.37$ was chosen again because the apparent FOT behavior 
is observed in $T < 0.3 T_{c0}$ \cite{Bianchi}. Further, we have assumed, 
for brevity, $T/T_{c0}=0.1$ and $\gamma=0.25$ (see eq.(\ref{eq:GL2})). Then, 
the 
$|\beta|$-value is located in the range between $2.0$ and $3.0$ used 
in our computation, and we have $\alpha_0$ ($\propto (N(0) T_{c0}/T)^{2/3}$) 
$\simeq 50$. This $\alpha_0$ value suggests that, on a linear scale of the 
magnetic field of the order of 
$H_{c2}(T=0)$, even the broader simulation curve 
in $|\beta|=2$ in Fig.\ref{fig:psi2} cannot be 
distinguished from a strictly discontinuous jump. 
Further, according to our MF calculation the (apparent) magnetization jump 
value at $T=50$(mK) is roughly estimated as $20$(G) which is comparable with 
the measured one ($\sim 30$(G)) \cite{Tayama}.

 In conclusion, we have given convincing results implying that 
a MF-FOT at $H_{c2}$ in superconductors with field-induced vortices 
(i.e., orbital depairing) should not occur as a genuine FOT in real systems 
with any nonvanishing fluctuation effect. 
Typically, this situation is realized in materials with a singlet pairing 
and with an enhanced paramagnetic depairing. Assuming a quasi 2D material with 
the $d_{x^2-y^2}$-pairing state in ${\mib H} \parallel c$, we have performed 
numerical simulations under 
a condition presumably compatible with high field CeCoIn$_5$ data suggesting a 
strong discontinuous transition at $H_{c2}$. 
The absence of a genuine FOT at $H_{c2}$ is consistent with the {\it absence} 
of hysteresis accompanying this discontinuous behavior in CeCoIn$_5$ 
and unrelated to the vortex pinning. 
The obtained MF phase diagrams seem to be roughly consistent 
with the observations in CeCoIn$_5$, including recent data \cite{Movsho} 
suggestive of a structural transition possibly to an FFLO-like vortex solid, 
if taking account of a nonvanishing impurity concentration. The issues 
in ${\mib H} \perp c$ will be considered elsewhere. 

After writing an earlier version of the manuscript, additional simulations 
at lower $T/T_{c0}$ ($< 0.1$) have been performed, and a small hysteresis 
in the almost discontinuous behaviors was found to occur in this case. 
However, since the GL coefficients (i.e., $V_4$ and $V_6$ in eq.(1)) remain 
unchanged, this hysteresis is {\it not} due to a genuine FOT but a phenomenon 
appearing on {\it measurable} Monte Carlo steps (i.e. time scales) as a 
reflection of incomplete relaxation in simulations \cite{Ada} 
which are inevitably in non-equilibrium. The tiny hysteresis appeared at 
lower $T$ but {\it above} $T_{\rm FFLO}$ in CeCoIn$_5$ only in 
${\mib H} \perp c$ \cite{Movsho} is likely to have a similar origin to 
this numerical one {\it practically} occurring in cases with sufficiently 
weak fluctuations. 

We thank T. Sakakibara, Y. Matsuda, K.Izawa, T. Tayama, R. Movshovich, 
and K. Machida for stimulative discussions. 
Part of the numerical computation in this work has been carried out at the 
Supercomputer Center, Institute for Solid State Physics, University of Tokyo, 
and at the Yukawa Institute Computer Facility in Kyoto University. 


\begin{thebibliography}{99}
\bibitem{Brandt} E. H. Brandt: Rep. Prog. Phys. {\bf 58} (1995) 1465.
\bibitem{RIr} R. Ikeda: J. Phys. Soc. Jpn. {\bf 70} (2001) 219.
\bibitem{Mineev} V. P. Mineev: Phil. Mag. B {\bf 80} (2000) 307.
\bibitem{Houzet} M. Houzet and A. Buzdin: Phys. Rev. B {\bf 63} (2001) 184521.
\bibitem{Izawa} K. Izawa {\it et al.}: Phys. Rev. Lett. {\bf 87} (2001) 057002.
\bibitem{Tayama} T. Tayama {\it et al.}: Phys. Rev. B {\bf 65} (2002) 180504.
\bibitem{Murphy} T. P. Murphy {\it et al.}: Phys. Rev. B {\bf 65} (2002) 100514(R).
\bibitem{Bianchi} A. Bianchi {\it et al.}: Phys. Rev. Lett. {\bf 89} (2002) 137002.
\bibitem{ISSP} A. Harita {\it et al.}: in Autumn Meeting of the Physical 
Society of Japan (2002). 
\bibitem{Bi} T. Shibauchi {\it et al.}: Phys. Rev. Lett. {\bf 86} (2001) 5763. 
\bibitem{Naughton} T. J. Lee {\it et al.}: Phys. Rev. Lett. {\bf 78} (1997) 3555.
\bibitem{Ada} H. Adachi and R. Ikeda: in preparation. 
\bibitem{Shima} H. Shimahara and D. Rainer: J. Phys. Soc. Jpn. {\bf 66} (1997) 3591. 
\bibitem{Moore} R. Ikeda {\it et al.}: J. Phys. Soc. Jpn. {\bf 61} (1992) 254. 
\bibitem{Ikeda3} R. Ikeda {\it et al.}: J. Phys. Soc. Jpn. {\bf 59} (1990) 1397. 
\bibitem{Tesanovic} Z. Tesanovic and L. Xing: Phys. Rev. Lett. {\bf 67} (1992) 2729. 
\bibitem{Kato} Y. Kato and N. Nagaosa: Phys. Rev. B {\bf 48} (1993) 7383. 
\bibitem{SI} S. Ikeda {\it et al.}: J. Phys. Soc. Jpn. {\bf 70} (2001) 2248. 
\bibitem{Suzuki} T. Suzuki {\it et al.}: J. Phys. Soc. Jpn. {\bf 69} (2000) 1462. 
\bibitem{03} A. A. Gapud {\it et al.}: Phys. Rev. B {\bf 67} (2003) 104516. 
\bibitem{Movsho} A. Bianchi {\it et al.}: cond-mat/0304420. 

\end{thebibliography}
\end{document}